\documentclass[prd,aps, preprint,10pt,nofootinbib,preprintnumbers,superscriptaddress]{revtex4-1}
\pdfoutput=1
\usepackage{geometry}                
\usepackage{graphicx}
\usepackage{tikz-feynman}
\usepackage{amssymb}
\usepackage{mathtools}
\usepackage{epstopdf}
\usepackage{verbatim}
\usepackage{color}
\usepackage{slashed}
\usepackage{bbold}
\usepackage{fullpage}
\usepackage{float}
\usepackage[title]{appendix}
\usepackage{amsfonts}
\usepackage{amssymb}
\usepackage{amsmath}
\usepackage{setspace}
\usepackage{epsfig}
\usepackage{hyperref}

\usepackage{amsthm}
\usepackage{tabularx}
\usepackage{empheq}
\usepackage{mathrsfs}
\usepackage[normalem]{ulem}

\definecolor{light-gray}{gray}{0.9}

\newcommand{\x}{\chi}
\newcommand{\Ap}{A^\prime}
\newcommand{\mAp}{m_{A^\prime}}

\newcommand{\eps}{\epsilon}

\newcommand{\order}[1]{\mathcal{O}{(#1)}}

\newcommand{\be}{\begin{eqnarray}}
\newcommand{\ee}{\end{eqnarray}}
\newcommand{\nn}{\nonumber}
\newcommand{\Eq}[1]{Eq.~(\ref{eq:#1})}

\newcommand{\Lagr}{\mathcal{L}}
\newcommand{\bea}{\begin{align}}
\newcommand{\eea}{\end{align}}
\newcommand{\bmat}{\begin{pmatrix}}
\newcommand{\emat}{\end{pmatrix}}
\newcommand{\bal}{\begin{align}}
\newcommand{\eal}{\end{align}}

\newcommand{\qeff}{q_\text{eff}}

\definecolor{colorRTD}{rgb}{.2,.2,.7}

\definecolor{colorJR}{rgb}{.3,.6,.1}

\newcommand{\w}{\omega}

\newcommand{\fDM }{f_{_{\text{DM}}}}

\newcommand{\vect}[1]{\boldsymbol{#1}}

\newcommand{\xv}{{\bf x}}
\newcommand{\vv}{{\bf v}}
\newcommand{\Vv}{{\bf V}}

\newcommand{\Ev}{\boldsymbol{E}}
\newcommand{\jv}{\boldsymbol{j}}
\newcommand{\zhat}{\hat{\boldsymbol{z}}}
\newcommand{\phihat}{\hat{\boldsymbol{\phi \, }}}

\newcommand{\grad}{\nabla}

\definecolor{myblue}{rgb}{.95, .95, 1}

\begin{document}

\title{Signals of Millicharged Dark Matter in Light-Shining-Through-Wall Experiments}

\author{Asher~Berlin}
\affiliation{Theoretical Physics Division, Fermi National Accelerator Laboratory, Batavia, IL 60510, USA}
\affiliation{Superconducting Quantum Materials and Systems Center (SQMS), Fermi National Accelerator Laboratory, Batavia, IL 60510, USA}
\author{Raffaele~Tito~D'Agnolo}
\affiliation{Universit\'e Paris-Saclay, CNRS, CEA, Institut de Physique Th\'eorique, 91191, Gif-sur-Yvette, France}
\author{Sebastian~A.~R.~Ellis}
\affiliation{D\'epartement de Physique Th\'eorique, Universit\'e de Gen\`eve, 
24 quai Ernest Ansermet, 1211 Gen\`eve 4, Switzerland}
\author{Jury~I.~Radkovski}
\affiliation{Department of Physics and Astronomy, McMaster University, 1280 Main Street West, Hamilton, ON L8S 4M1, Canada}
\affiliation{Universit\'e Paris-Saclay, CNRS, CEA, Institut de Physique Th\'eorique, 91191, Gif-sur-Yvette, France}

\affiliation{Perimeter Institute for Theoretical Physics, Waterloo, Ontario, N2L 2Y5, Canada}

\begin{abstract}

We discuss a novel detection technique for millicharged dark matter that makes use of existing light-shining-through-wall (LSW) experiments searching for massive dark photons. Since millicharged particles interact with both the visible and dark sectors, a small background of such particles enables the search for visible signals even in the limit of a massless dark photon. Furthermore, for sufficiently large couplings, a small dark matter subcomponent consisting of millicharged particles can have a terrestrial abundance that is significantly enhanced with respect to its galactic abundance. We leverage this fact to show that in certain parts of parameter space, future runs of the LSW Dark SRF experiment can be used to set the strongest limits on millicharged relics.

\end{abstract}
\maketitle

\section{Introduction}

The mystery that is the nature of dark matter (DM) remains unsolved despite many decades of theoretical and experimental advances. An intriguing possibility is that DM could be at least partially composed of particles with extremely small fractional charges under electromagnetism, commonly referred to as ``millicharged particles" (mCPs). Although no mCPs have been observed, there is strong motivation for their existence, as they naturally arise  in various string theory compactifications and minimal dark sector extensions to the Standard Model (SM) involving a gauged $U(1)^\prime$ symmetry~\cite{Holdom:1985ag, Dienes:1996zr, Abel:2003ue, Batell:2005wa, Aldazabal:2000sa, Abel:2004rp,Abel:2008ai,Acharya:2016fge,Acharya:2017kfi,Gherghetta:2019coi} and could also play a role in explaining recent experimental anomalies~\cite{PVLAS:2005sku,PAMELA:2008gwm,Chang:2008aa,Barkana:2018lgd,Bowman:2018yin,Berlin:2018sjs,Barkana:2018qrx,Liu:2019knx}. The boson of this $U(1)^\prime$, referred to as an $\Ap$ or dark photon, may possess a kinetic mixing with the SM photon, parameterized by a small dimensionless parameter $\epsilon$. In simple field-theoretic models, this arises from loop diagrams involving fields charged under both the SM and dark sectors, leading to mixings of $\epsilon \sim 10^{-6} - 10^{-3}$~\cite{Holdom:1985ag}, but much smaller mixings are possible and have been explicitly constructed~\cite{Arkani-Hamed:2008kxc,Abel:2008ai,Acharya:2017kfi,Gherghetta:2019coi}. As a result, particles directly charged in the dark sector under the $\Ap$ appear as effectively millicharged under normal electromagnetism. 

Many experiments have been conducted in an attempt to detect mCPs with effective charge greater than $\sim 10^{-4}$~\cite{Agrawal:2021dbo}. These have all turned up null results, leading to a number of proposals to search for mCPs with even smaller charges and heavier masses~\cite{Prinz:1998ua,Davidson:2000hf,Battaglieri:2017aum,Berlin:2018bsc,Magill:2018tbb,Kelly:2018brz,Chang:2018rso,Harnik:2019zee,ArgoNeuT:2019ckq,Ball:2020dnx,Budker:2021quh}. Despite being very feebly-coupled to normal matter, the long range of the interaction enhances the mCP-SM scattering cross-section in the non-relativistic limit. As a result, throughout much of the parameter space, a millicharged DM (mCDM) subcomponent rapidly thermalizes with the Earth's environment, cooling down to terrestrial temperatures of $300 \ \text{K} \sim 25 \ \text{meV}$~\cite{Pospelov:2020ktu,Berlin:2023zpn}. Terrestrial direct detection experiments with $\text{eV} - \text{keV}$ energy thresholds are thus largely insensitive to such DM subcomponents, motivating the consideration of other approaches~\cite{Emken:2019tni,Pospelov:2020ktu,Berlin:2023zpn,Budker:2021quh,Berlin:2021zbv,McKeen:2022poo,Billard:2022cqd,Das:2022srn}. This shedding of kinetic energy also implies a drastic modification to the mCDM phase space; analogous to a ``traffic jam," conservation of flux implies that the terrestrial density of thermalized mCDM is drastically enhanced compared to its average galactic density~\cite{Wallemacq:2013hsa,Wallemacq:2014lba,Wallemacq:2014sta,Neufeld:2018slx,Laletin:2019qca,Pospelov:2019vuf,Pospelov:2020ktu,Berlin:2023zpn,Leane:2022hkk}. 

\begin{figure}
\includegraphics[width = 0.7 \textwidth]{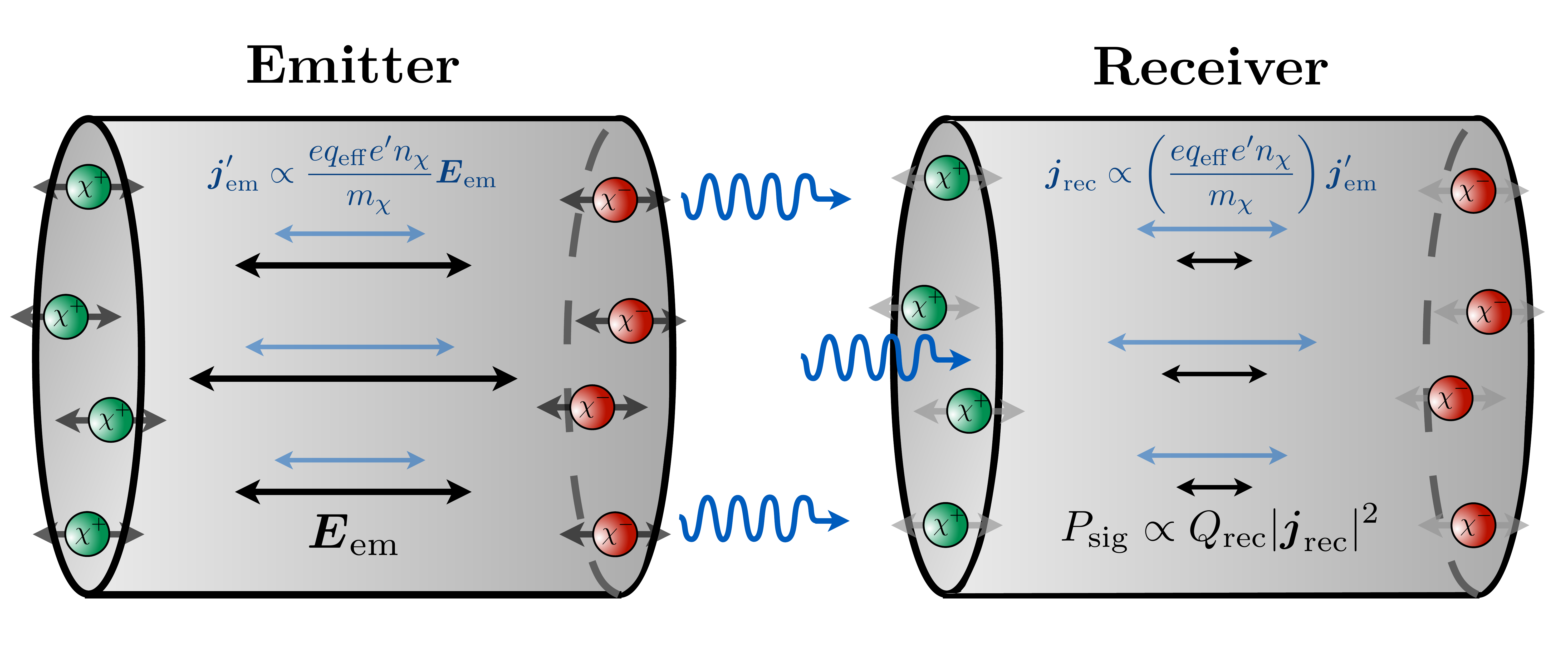}
\caption{Cartoon of an LSW experiment with an mCP-induced signal. An emitter cavity is loaded with EM energy in a resonant mode, which causes mCPs in the cavity to oscillate. The mCP oscillations generate a small dark current $\vect{j}'_{\rm em}$ as given in \Eq{jem1}. The invisible radiation sourced by this current excites the same mode in a shielded receiver cavity (\Eq{jrecGen}), leading to a resonantly-enhanced signal power (\Eq{PsigGen}).
}
\label{fig:Cartoon}
\end{figure}

The large density and small kinetic energy of such particles makes them ideal candidates to search for in \emph{direct deflection} experiments~\cite{Berlin:2019uco,Berlin:2021kcm}. As first pointed out in Ref.~\cite{Berlin:2019uco}, it is possible to induce collective effects in the background fluid of particle-like DM, which can be leveraged to improve detection prospects in the low momentum-transfer regime. In this work, we investigate mCDM signals arising from such collective effects in light-shining-through-wall (LSW) experiments. As shown in Fig.~\ref{fig:Cartoon}, such a setup consists of two identical radio-frequency (RF) cavities separated by electromagnetic (EM) shielding. One cavity is loaded in a resonant mode at frequency $\omega \sim 1 \ \text{GHz}$. Inside this emitter cavity, the resulting electric field generates a perturbation in the mCDM density that sources \emph{dark} electric fields capable of penetrating through the shielding. The dark electric fields couple to mCPs inside a second detector cavity, causing them to oscillate and source \emph{visible} electric fields that are then resonantly detected. This effect relies on the presence of a kinetically-mixed $\Ap$, but unlike direct LSW searches for dark photons, does not decouple in the massless $\Ap$ limit. Although massless dark photons do not directly couple to the SM in vacuum, mCPs interact with both sectors, giving rise to rich phenomenology even in the case that the $U(1)^\prime$ symmetry remains unbroken~\cite{Berlin:2022hmt,Vogel:2013raa,Adshead:2022ovo}. Furthermore, we find that for coaxially-aligned cavities in the far-field limit, it is ideal to employ transverse electric modes, which is suboptimal for typical setups searching for signals arising from massive dark photons in absence of a mCDM background~\cite{Betz:2013dza,Graham:2014sha,Romanenko:2023irv,Berlin:2022hfx}. Regardless, it is still possible to reinterpret existing limits from experiments performed at CERN and FNAL~\cite{Betz:2013dza,Romanenko:2023irv} to place new bounds on mCDM. In this work, we also compute the projected reach of a future experiment dedicated to enhancing the sensitivity to mCDM. Our main results are summarized in Figs.~\ref{fig:ReachPlot1e3} and \ref{fig:ReachPlotNX}, which show that future runs of the Dark SRF experiment~\cite{Romanenko:2023irv} can probe orders of magnitude of new parameter space.

The remainder of this study is organized as follows. In Section~\ref{sec:milliQ}, we give a brief overview of mCDM models coupled to massless dark photons. In Section~\ref{sec:setup}, we discuss mCDM signals in LSW experiments. Section~\ref{sec:results} presents existing limits and the estimated sensitivity of current and future LSW experiments. We conclude in Section~\ref{sec:conclusion}. 

\section{Model Overview of Millicharged Particles}\label{sec:milliQ}

We consider a DM subcomponent consisting of mCPs (denoted as $\x$), with an effective charge of $\qeff \ll 1$ under normal electromagnetism. In this section, we give an overview for how such interactions can arise. If electric charge is quantized, as expected if $U(1)_{\rm EM}$ originates from the breaking of a Grand Unified Theory, the simplest realization of mCPs arises from the kinetic mixing between $U(1)_{\rm EM}$ and an unbroken dark $U(1)'$. The Lagrangian of the theory is then
\begin{align}
\label{eq:Lag1}
\Lagr \supset -\frac{1}{4} \,  F_{\mu\nu}F^{\mu\nu} - \frac{1}{4} \, F'_{\mu\nu} \, F'^{\mu\nu} + \frac{\epsilon}{2} \, F^{\mu\nu} \, F'_{\mu\nu} - e \, A_\mu \, j^\mu - e' \, A'_\mu \, j'^\mu
~,
\end{align} 
where $A^{(\prime) \, \mu}$ are the gauge fields, $F_{\mu \nu}^{(\prime)}$ the field strengths, $e^{(\prime)}$ the gauge couplings, and $j^{(\prime) \, \mu}$ the current densities, such that a prime denotes a dark sector quantity. The signals discussed in this work dominate over existing LSW searches for massive dark photons (see, e.g., Ref.~\cite{Caputo:2021eaa} and references within) in the limit that the mass is $\mAp \ll 1/L_{\rm exp}$, where $L_{\rm exp}$ is the typical size of the experimental apparatus. Therefore, here and in the following we take the $\Ap$ to be massless. The transformation $A'^\mu \to A'^\mu + \eps A^\mu$ brings the kinetic terms into a canonical form to leading order in $\epsilon \ll 1$, such that in this new basis \Eq{Lag1} becomes
\begin{align}
\label{eq:mCPbasis}
\Lagr \supset -\frac{1}{4} \,  F_{\mu\nu}F^{\mu\nu} - \frac{1}{4} \, F'_{\mu\nu} \, F'^{\mu\nu} - A_\mu \, \big( e \, j^\mu + \eps \, e^\prime \, j'^\mu \big) - e' \, A'_\mu \, j'^\mu
~.
\end{align} 
Note that the dark photon field solely couples to the dark sector current. In other words, massless dark photons do not directly couple to SM sources. As a result, typical LSW searches are usually thought to lack sensitivity to a massless $\Ap$. However, this is not the case if mCPs are present. This can be seen by noting that the SM photon couples to both the SM and dark sector current densities, with the latter interaction suppressed by $\eps$. Thus, dark sector particles $\x$ directly charged under the $\Ap$ couple to the SM photon with an effective charge 
\be
\qeff \equiv \epsilon \, e' / e\, . 
\ee
Since dark currents couple to both sectors, the invisible mode \emph{indirectly} couples to SM currents if it first interacts with an EM-charged mCP background. 

This decoupling in the massless limit can also be understood by phrasing the choice of basis in terms of the $(A, A^\prime)$ plane and the $\text{SO}(2)$ symmetry of the kinetic terms. Absent of mCPs ($j'^\mu = 0$), \Eq{Lag1} implies that the coupling to $j^\mu$ identifies a preferred direction in the $(A, A^\prime)$ plane, but since the rest of the Lagrangian is $\text{SO}(2)$ symmetric, there is no physical meaning to this specific direction; we can rotate it at will without changing the rest of the Lagrangian. An observer will call ``the SM photon" any linear combination of fields that couples to $j^\mu$. This is why the massless invisible mode is unobservable in vacuum. Alternatively, in this language it is simple to see that the massless invisible mode is instead observable in the presence of mCPs ($j'^\mu \neq 0$). In this case, there are two preferred directions in the $(A, A^\prime)$ plane, given by the linear combinations of fields that couple to $j^\mu$ or $j'^\mu$. A rotation in the $(A, A^\prime)$ plane does not change the scalar product between these two directions, so one can calculate basis-independent quantities such as the degree to which the linear combination of fields that couples to $j^\mu$ also couples to $j'^\mu$. Hence, for $j'^\mu = 0$ we have access only to a single vector and will never know if it is embedded in a plane. But if $j'^\mu \neq 0$, we can perform measurements that reveal the existence of a second vector in the $(A, A^\prime)$ plane and thus of the plane itself. The same reasoning can also be applied to the well-studied example of a massive dark photon absent of mCPs; in this case, the massive $\Ap$ is observable even if $j'^\mu = 0$ because $j^\mu$ and $\mAp$ specify two preferred directions in the $(A, A^\prime)$ plane.

The lightest dark sector particles $\x$ directly charged under the $\Ap$ are absolutely stable and are therefore natural DM candidates. In the simplest of cosmologies, such particles arise as DM subcomponents in the form of thermal relics. For effective charges larger than $\qeff \gtrsim 10^{-7} \times (m_\x / \text{GeV})^{1/2}$, electron annihilations $e e \to \x \x$ thermalize mCPs with the SM sector in the early universe, where $m_\x$ is the mCP mass. Such mCPs generically constitute a small fraction $\fDM$ of the total DM energy density, since freeze-out through $\x \x \to \Ap \Ap$ easily depletes the thermal mCP abundance to $\fDM \sim 10^{-8} \times (m_\x / \text{GeV})^2 \, (1 / e^\prime)^4$. Such cosmologies in which the dark sector efficiently equilibrates with the SM at a temperature below $\sim 1 \ \text{GeV}$ predict a sizable relativistic density of dark photons, which is tightly constrained by early universe probes of additional dark radiation~\cite{Vogel:2013raa,Adshead:2022ovo}. Such bounds are weakened for heavier mCP masses, since the two sectors decouple at temperatures $T \sim m_\x$; if this occurs at early enough times, the multiplicity of SM mass thresholds heats the SM relative to the dark sector, decreasing the relative density in dark radiation. These bounds are also evaded in cosmological scenarios predicting exponentially small values of $\fDM$, where the reheat temperature of the universe $T_\text{RH}$ is significantly smaller than the mass $m_\x$. In this case, the dark sector never fully thermalizes with the SM in the early universe, yielding $\fDM \sim (\alpha_\text{em} \, \qeff)^2 \, e^{-2 m_\x / T_\text{RH}} \,  m_\chi m_\text{pl} / (T_\text{RH} \, T_\text{eq}) \ll 1$~\cite{Berlin:2021zbv}, where $m_\text{pl}$ is the Planck mass and $T_\text{eq} \sim 1 \ \text{eV}$ is the temperature at matter-radiation equality.

As a reference point for the experimental signals considered in this work, we can consider a model of mCDM with $m_\chi \gtrsim \text{few} \times \text{MeV}$, where $\chi$ is a small fraction of DM, $\fDM\ll 1$. For such masses, we can invoke a small reheat temperature $T_\text{RH}$ to evade cosmological bounds on dark radiation. Small mCDM subcomponents are observable in the laboratory today because interactions with terrestrial matter can significantly enhance their local density $n_\x^\oplus$ compared to their mean number density throughout the galaxy $n_\x^{(\text{gal})} \sim \order{1} \ \text{cm}^{-3} \times \fDM \, (\text{GeV} / m_\x)$. Specifically, for $\qeff \gg 10^{-7} \times (m_\x / \text{GeV})^{1/2}$, galactic mCPs bombarding the Earth rapidly equilibrate to terrestrial temperatures after scattering with nuclei in the atmosphere and crust. Various dynamics tied to this process can give rise to large local overdensities. For instance, ``strongly-coupled'' mCPs much heavier than $\sim 1 \ \text{GeV}$ accumulate on Earth over geological timescales, since their characteristic thermal velocity is well below the terrestrial escape velocity. More generally, the large collision rate with normal matter increases the residence time spent by mCPs. Therefore, as such particles diffuse past the surface, eventually escaping the Earth entirely or settling into hydrostatic equilibrium elsewhere, conservation of flux implies a local enhancement to their density. Such effects have been studied in Refs.~\cite{Wallemacq:2013hsa,Wallemacq:2014lba,Wallemacq:2014sta,Neufeld:2018slx,Laletin:2019qca,Pospelov:2019vuf,Pospelov:2020ktu,Leane:2022hkk,Berlin:2023zpn}, which have showed that the terrestrial mCDM density can be as large as $n_\x^\oplus \sim \order{10^{15}} \ \text{cm}^{-3} \times \fDM$.

A dedicated analysis of the formation of terrestrial mCP densities (required to map $\fDM$ onto $n_\x^\oplus$) is beyond the scope of this work. We calculate limits and projections on the \emph{laboratory} density $n_\x$ of mCPs instead of $\fDM$, remaining agnostic regarding the formation of such a population and possible modifications from mCP self-interactions. Beyond simplifying the study, there are various reasons to remain agnostic. First, we will focus exclusively on mCP densities many orders of magnitude smaller than the density of normal matter, in which case the timescale associated with self-interactions is often longer than the time it takes for such terrestrial densities to develop~\cite{Pospelov:2020ktu}. Furthermore, processes in the early universe or at late times may lead to a sizable mCP asymmetry, such that some of these interactions are forbidden. For instance, for $\qeff \gtrsim m_e / \min{(m_\x , m_N)}$ negatively-charged mCPs efficiently form bound states with atomic nuclei $N$, such that the freely propagating mCP population on Earth consists dominantly of positively-charged particles~\cite{Pospelov:2019vuf,Berlin:2021zbv}. The signal we discuss here remains observable even in the case of an asymmetric population of mCPs.

\section{Light Shining Through a Wall}\label{sec:setup}

LSW experiments searching for direct signals of dark photons include ALPS~\cite{Ehret:2010mh,Ortiz:2020tgs,Hallal:2020ibe}, CROWS~\cite{Betz:2013dza}, and Dark SRF~\cite{Romanenko:2023irv}. In this work, we show that these same searches are also sensitive to mCPs, and, furthermore, that this sensitivity can be enhanced with small modifications to the experimental geometry. We solely focus on LSW setups employing RF cavities, since this technology has been firmly established by the CROWS experiment, and will be developed to a much greater degree by the Dark SRF collaboration in the coming years. 

\subsection{Review of Massive Dark Photon Signals}

CROWS and Dark SRF search for massive dark photons by first driving fields of amplitude $E_\text{em} \sim 10 \ \text{MV} / \text{m}$ and frequency $\w \sim 1 \ \text{GHz}$ in an emitter cavity. The SM photons source a dark photon field, which travels unimpeded into a shielded receiver cavity, exciting a small visible EM field of the same frequency. In Ref.~\cite{Graham:2014sha}, it was pointed out that cavities driven in the $\text{TM}_{010}$ mode in a longitudinal arrangement (where the cavities are aligned along the polarization axis of the electric field) have optimal sensitivity to ultralight dark photons. 

\begin{figure}[h]
    \begin{tikzpicture}
\begin{feynman}
\vertex (m1);
\node[right=1.5cm of m1] (m2) [crossed dot];
\vertex[right=1.5cm of m2] (m3);
\vertex[above left=1.6cm of m1] (u1);
\vertex[below left=1.6cm of m1] (d1);
\diagram* {
(u1) -- [fermion] (m1) -- [fermion] (d1),
(m1) -- [photon, edge label=$A$] (m2) ,
(m2) -- [boson, thick, edge label=$A'$, near end] (m3),
};
\vertex [above=1.1em of m2] {$\epsilon \, m^2_{A'}$};
\vertex [left=1cm of m1] {$j_{\mu}$};
\end{feynman}
\end{tikzpicture}
    \caption{Leading diagram for the emission of a massive kinetically-mixed dark photon $A^\prime$ from a SM electromagnetic current $j_\mu$.}
     \label{fig:FeynmanDiagram}
\end{figure}
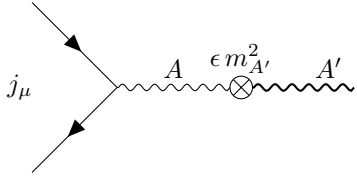

This result will be useful also in our setup. To understand it we can follow Ref.~\cite{An:2013yfc} and rewrite the interaction between the dark photon and the ordinary photon in the original gauge basis such that
\be
\mathcal{L}=-\frac{1}{4} \,  F_{\mu\nu}F^{\mu\nu} - \frac{1}{4} \, F'_{\mu\nu} \, F'^{\mu\nu} - e A_\mu \, \, j^\mu - \epsilon \, \mAp^2 \, A'_\mu \, A^\mu\, .
\ee
Dark photons are emitted as in Fig.~\ref{fig:FeynmanDiagram}, with an amplitude
\be
\mathcal{M}_{i\to f+ A^\prime}= \epsilon \, e \, \mAp^2 \, (j_\mu)_{if} \, \langle A^\mu, A^\nu\rangle \, \epsilon_\nu^\prime
\, .
\ee

Here, $(j_\mu)_{if}$ is the current matrix element between initial and final states, $\langle A^\mu, A^\nu\rangle$ is the photon propagator in the cavity, and $\epsilon_\nu^\prime$ the emitted dark photon polarization. The cavity walls break translational invariace, giving the photon an effective mass that corresponds to the normal modes (labelled by $n$) of the EM field in the cavity. In Coulomb gauge ($\nabla \cdot \mathbf{A}=0$), we have 
\be
\langle A_i, A_j\rangle =\sum_{n=0}^{\infty} \frac{\delta_{ij}}{\w^2-\w_n^2} \nn
~~,~~
\langle A_0, A_0\rangle = 0\, , \label{eq:prop}
\ee
where $\w_n$ are the wavenumbers of the normal modes of the cavity (eigenvalues of $\nabla^2$).
Using the well-known expressions of $\epsilon^\prime_\nu$ for a massive vector, we can conclude that in the limit of interest ($\mAp \ll 1/L_{\rm cav}$), longitudinal modes $A_L^\prime$ are emitted with a higher rate than transverse ones $A_T^\prime$,
\be
\Gamma_{i\to f+A^\prime_L} \propto \epsilon^2 \, \frac{\mAp^2}{\w^2}\, , \quad 
\Gamma_{i\to f+A^\prime_T} \propto \epsilon^2 \, \frac{\mAp^4}{\w^4}\, , \label{eq:massive}
\ee
where our expressions are valid for $\w > \w_0$, such that $\w_0$ is the lowest-lying mode of the cavity $\w_0 \sim 1/L_{\rm cav}$. This reverses the intuition of emission in vacuum where $\Gamma_{i\to f+A^\prime_L}\propto \epsilon^2 \mAp^2/\w^2$ and $\Gamma_{i\to f+A^\prime_T} \propto \epsilon^2$. Thus, in the limit that $\mAp \ll \w \sim 1 \ \mu \text{eV}$, the longitudinal LSW arrangement has optimal sensitivity to massive dark photons. On resonance ($\w\simeq \w_0$),  the finite widths of the modes need to be to included in \Eq{prop}, $\langle A_i, A_j\rangle \sim \delta_{ij} (\w^2-\w_0^2-i \w \w_0/Q_0)^{-1}$, but this does not modify the overall scalings of $\Gamma_{i\to f+A^\prime_L}\sim \epsilon^2 \mAp^2$ and $\Gamma_{i\to f+A^\prime_T}\sim \epsilon^2 \mAp^4$.

\subsection{Massless Dark Photons in a Millicharged Plasma}\label{sec:signal}

Similar signals arise for massless dark photons provided that there is an ambient density of mCPs, even though there is no mass splitting between the visible and dark transverse modes. Qualitatively, this originates from the plasma mass of the dark photon induced by the mCP density in the cavities. However, this does not constitute a physical mass and care must be taken when recasting the reach of LSW experiments to mCPs. For simplicity, we will take the mCDM to be equally dense in the emitter and receiver cavities. In principle, this need not be the case if the two cavities are kept in separate environments with different temperatures, which can alter the local mCDM density in either location. More generally, our adopted values of $n_\x$ should be interpreted as the geometric mean of the mCP density in the two cavities.

Before providing a detailed derivation of such signals, we first give a heuristic discussion of the effect, working in the basis of \Eq{mCPbasis} in which $\x$ couples to both visible and invisible fields. In the emitter cavity, mCPs are accelerated by the driven electric field $E_\text{em}$, setting up a dark current oscillating at the same frequency with amplitude $j_\text{em}^\prime \sim (e^\prime n_\x / m_\x) \, (e q_\x  E_\text{em} / \w)$. This dark current sources a dark EM field of $E_\text{em}^\prime \sim j_\text{em}^\prime / \w$, which penetrates into the receiver cavity, accelerating the mCPs contained there and inducing a visible current of $j_\text{rec} \sim (e q_\x n_\x / m_\x) \, (e^\prime E_\text{em}^\prime / \w)$. This visible current thus resonantly excites the receiver mode to a level of
\be
\label{eq:EsigParametric}
E_\text{sig} \sim Q \,  j_\text{rec} / \w \sim Q \, \eps^2  \,  (\w_p^\prime / \w)^4 \, E_\text{em}
~,
\ee
where 
\be
\w_p^\prime \equiv \sqrt{e^{\prime \, 2} \, n_\x  / m_\x} \label{eq:wpd}
\ee
is the in-medium dark plasma frequency as sourced by the mCP background. Comparing \Eq{EsigParametric} to the signal from massive dark photons in \Eq{massive}, we see that the mCP-induced signal scales similarly to setups measuring the \emph{transverse} mode of a massive $\Ap$, with the dark plasma frequency $\w_p^\prime$ playing the role of an effective dark photon mass. We now substantiate this estimate with a detailed calculation.

An effective coarse-grained description of a dense mCP background is that of a fluid. The equations of motion of this mCP fluid can be derived starting from the non-relativistic Boltzmann equation, 
\be
\label{eq:Boltz1}
\frac{d f_\pm}{d t} \simeq \frac{1}{m_\x} \, C [f_\pm]
~,
\ee
where $f_\pm$ is the phase space density for positively- or negatively-charged mCPs and $C [f_\pm]$ is the collision operator. The LHS of \Eq{Boltz1} models the evolution of phase space elements arising from density/temperature gradients and background fields, whereas the RHS incorporates additional effects from, e.g., collisions between mCPs and nuclei or other mCPs.  In the presence of background visible or invisible electric fields $\Ev$ and $\Ev^\prime$, the time-derivative in \Eq{Boltz1} can be expanded as
\be
\label{eq:chainrule}
\frac{d f_\pm}{d t} \simeq \frac{\partial f_\pm}{\partial t} + \vv_\pm \cdot \frac{\partial f_\pm}{\partial \xv} \pm \frac{1}{m_\x} \, \big(e \qeff \, \Ev +  e^\prime \Ev^\prime \big) \cdot \frac{\partial f_\pm}{\partial \vv_\pm}
~,
\ee
where $\vv_\pm$ is the $\x^\pm$ particle velocity. In  \Eq{chainrule}, we have ignored magnetic fields since their effect is suppressed in the non-relativistic limit. Using this in Eq.~({\ref{eq:Boltz1}) and taking the first moment over the mCP phase space yields
\be
\label{eq:Boltz2}
\partial_t \Vv_\pm + (\Vv_\pm \cdot \nabla) \, \Vv_\pm + \frac{\nabla P_\pm}{ n_\pm \, m_\x} \simeq \pm \frac{1}{m_\x} \, \big(e \qeff \, \Ev +  e^\prime \Ev^\prime \big) + \frac{m_\x^2}{n_\pm} \int \frac{d^3 \vv_\pm}{(2 \pi)^3} ~ \vv_\pm \, C [f_\pm]
~,
\ee
where $n_\pm$, $\Vv_\pm$, and $P_\pm$ are the number density, bulk velocity, and pressure of the positively- or negatively-charged mCP fluid. 
In general, the collision term $C[f_\pm]$ includes contributions from scattering processes involving same-charge and opposite-charge mCPs. However, when integrated over the velocity as in the last term of \Eq{Boltz2}, only scattering between oppositely-charged particles contributes, since interactions between identical particles do not change the overall bulk velocity $\Vv_\pm$ of either the positively- or negatively-charged fluid.

In evaluating the LHS of \Eq{Boltz2}, we have ignored terms arising from higher-order viscosity moments of the mCP phase space. This is justified in the limit that the mCP phase space is well-approximated by a boosted Maxwellian, whose only moments are $n_\x$ (the monopole) and $\Vv_\pm$ (the dipole). Indeed, since in the parameter space of interest the ambient matter efficiently thermalizes the mCPs before they enter the cavity, higher-order moments (e.g., quadrupole, octupole, etc.) are parametrically suppressed by the large mCP-atomic scattering rate~\cite{Dodelson:2003ft}. Under this same approximation and also taking the bulk velocity to be small compared to the thermal velocity ($V_\pm \sim e \qeff \, E / (m_\x \w) \ll \sqrt{T_\x / m_\x}$), the collision term on the RHS of \Eq{Boltz2} has been evaluated in Refs.~\cite{Dvorkin:2013cea,Boddy:2018wzy,Becker:2020hzj,Dvorkin:2020xga} within the context of DM-baryon scattering in the early Universe.\footnote{Note that for a receiver cavity cooled down to $T_\text{cav}  =10 \ \text{mK}$, the approximation $V_\pm\ll \sqrt{T_\x / m_\x}$ does not hold for sufficiently large couplings. In this case, the exact form of the collision term is modified compared to Eqs.~\eqref{eq:Boltz3} and \eqref{eq:Gamma1}. However, this does not affect our results since collisions have a negligible impact on the signal within the entirety of our parameter space, as discussed below \Eq{wplasma}.} These results can be easily adapted to incorporate $\x^+ - \x^-$ and $\x^\pm - \text{atomic}$ scattering to give
\be
\label{eq:Boltz3}
\partial_t \Vv_\pm + (\Vv_\pm \cdot \nabla) \, \Vv_\pm + \frac{\nabla P_\pm}{ n_\pm \, m_\x} \simeq \pm \frac{1}{m_\x} \, \big(e \qeff \, \Ev +  e^\prime \Ev^\prime \big) - \Gamma_\x \, (\Vv_\pm - \Vv_\mp) - \Gamma_\text{SM} \, (\Vv_\pm - \Vv_\text{SM})
~.
\ee
Above, $\Gamma_\x$ and $\Gamma_\text{SM}$ are the relevant momentum-exchange rates from mCP self-scattering and scattering with SM matter, respectively, 
\be
\label{eq:Gamma1}
\Gamma_i \simeq \frac{\mu_{\x i}}{3 m_\x \, \bar{v}_0^2} \, n_i \, \langle \sigma_T \, v_\text{rel}^3 \rangle
~,
\ee
where $i = \x$ or  $i = \text{SM}$, $v_\text{rel}$ is the relative velocity between $\x$ and the target scatterer, $\mu_{\x i}$ the $\x-i$ reduced mass, and $\sigma_T$ the momentum-transfer cross-section. The brackets correspond to a thermal average over $v_\text{rel}$ with the distribution
\be
f_\text{rel} (v_\text{rel}) = \frac{1}{(2 \pi)^{3/2} \, \bar{v}_0^3} \, e^{-v_\text{rel}^2 / 2 \bar{v}_0^2}
~~,~~
\bar{v}_0 \equiv \sqrt{T_\x / m_\x + T_i / m_i}
~,
\ee
with species $i$ at temperature $T_i$.

We reduce \Eq{Boltz3} further still by making several additional simplifying approximations. First, the low density of normal matter  inside the RF cavities implies that $\Gamma_\x \gg \Gamma_\text{SM}$.  Second, we drop the terms involving $(\Vv_\pm \cdot \nabla) \Vv_\pm$ and $\nabla P_\pm$ since both are higher-order in the non-relativistic (small temperature) limit. Also note that such gradient terms are additionally suppressed if the emitter cavity is driven in its lowest-lying transverse-magnetic mode, in which case the visible electric field $E$ is spatially uniform and is thus not expected to lead to significant spatial gradients in the mCP phase space. With these simplifications, \Eq{Boltz3} reduces to
\be
\label{eq:Boltz4}
\partial_t \Vv_\pm \simeq \pm \frac{1}{m_\x} \, \big(e \qeff \, \Ev +  e^\prime \Ev^\prime \big) - \Gamma_\x \, (\Vv_\pm - \Vv_\mp)~.
\ee

The electric fields in \Eq{Boltz4} are sourced by the oscillating driven modes of the emitter cavity. Taking these to be oscillating with frequency $\w$, \Eq{Boltz4} can be solved for the dark mCP current, 
\be
\label{eq:jx1}
\jv^\prime = e^\prime \, (n_+ \Vv_+ - n_- \Vv_-) \simeq - \frac{i}{\w} \, \frac{(e^\prime / e \qeff) \, \w_p^2 \, \Ev + \w_p^{\prime \, 2} \, \Ev^\prime}{1 - 2 i \, \Gamma_\x / \w}
~,
\ee
where we took $n_+ = n_- = n_{\chi}/2$ and defined the mCP contribution to the SM plasma frequency
\be
\w_p \equiv \sqrt{(e \qeff)^2 \, n_\x  / m_\x}
~. 
\label{eq:wplasma}
\ee
Note that \Eq{jx1} implies that $j^\prime$ is suppressed in the strong-coupling limit, $\Gamma_\x \gg \w$. Numerically, we find that this is not the case and that $\Gamma_\x/\w\ll1$ in all of the parameter space of interest. Thus, we set $\Gamma_\x \to 0$ in \Eq{jx1}. In this case, the dark electric field sourced by $j^\prime$ is $E^\prime \sim j^\prime / \w$. For sufficiently large dark coupling, this dark electric field can backreact on the mCPs, inhibiting growth of the current $j^\prime$. To see when this occurs, take the mCP current to be initially excited by the visible field of the emitter cavity, i.e., $j^\prime \sim (e^\prime / e \qeff) \, \w_p^2 \, E / \w$. Thus, the dark field sourced by this current is $E^\prime \sim (e^\prime / e \qeff) \, (\w_p / \w)^2 E$. From \Eq{jx1}, we see that this contribution dominates over that of the visible field if $\w_p^\prime \gtrsim \w$. Note that this agrees with the intuition that if the timescale for the mCP fluid to screen the emitter cavity, $\sim 1/\w_p^\prime$, is longer than the oscillation timescale $\sim 1/\w$ of the driven cavity, then backreactions can be ignored. We have checked that  $\w_p^\prime \lesssim \w$ for all of the parameter space of interest, in which case we can ignore such processes. 

\Eq{jx1} can also be used to determine the corresponding dark charge density $\rho^\prime$, which is related via continuity $\rho^\prime = (i/\w) \, \nabla \cdot \jv^\prime$. Hence, at the level of these approximations, the dark current and charge induced in the emitter cavity are 
\be
\label{eq:jem1}
\jv_\text{em}^\prime \simeq - \frac{i}{\w} \,  \Big( \frac{e \qeff  \, e^\prime \, n_\x}{m_\x} \, \Ev_\text{em} + \w_p^{\prime \, 2} \, \Ev_\text{em}^\prime \Big)
~~,~~
\rho_\text{em}^\prime \simeq  \frac{e \qeff  \, e^\prime \, n_\x}{m_\x} \, \frac{\nabla \cdot \Ev_\text{em}}{\w^2 - \w_p^{\prime \, 2}}
~,
\ee
where we used that $\grad \cdot \Ev_\text{em}^\prime = \rho_\text{em}^\prime$. Note that since $\nabla \cdot \Ev_\text{em} = \rho + \order{\eps}$ vanishes  in the interior vacuum of the cavity to leading order in $\eps \ll 1$, non-zero dark charge density $\rho_\text{em}^\prime$ can only be induced in the cavity walls. The dark charge and current densities in \Eq{jem1} source an oscillating invisible EM field emanating from the emitter cavity. In particular, using \Eq{jem1} in the standard form of the wave equation, $(\grad^2 + \w^2) \Ev_\text{em}^\prime = \partial_t \jv_\text{em}^\prime + \grad \rho_\text{em}^\prime$, this field is
\be
\label{eq:EinvGen}
\Ev_\text{em}^\prime (\xv, t) = - \frac{e \qeff \, e^\prime \, n_\x}{4 \pi \, m_\x} \, e^{i \w t} \int_\text{em} \hspace{-0.2 cm} d^3 \xv^\prime ~ \frac{e^{-i k |\xv - \xv^\prime|}}{|\xv - \xv^\prime|} \, \Big[\Ev_\text{em} (\xv^\prime) + \frac{1}{\w^2 - \w_p^{\prime \, 2}} \, \grad \grad \cdot \Ev_\text{em} (\xv^\prime) \Big]
~,
\ee
where the integral is performed over the volume of the emitter cavity and we defined the wavenumber
\be
k = 
\begin{cases}
\sqrt{\w^2 - \w_p^{\prime \, 2}} &(\w_p^\prime < \w)
\\
-i \, \sqrt{\w_p^{\prime \, 2} - \w^2} &(\w_p^\prime > \w)
~.
\end{cases}
\ee
Note that for $\w_p^\prime \gg \w$, $E_\text{em}^\prime$ is exponentially suppressed, as expected from the heuristic discussion above. 

This invisible field propagates unattenuated into the receiver cavity where it drives a corresponding \emph{visible} mCP current density, which is determined analogously to Eqs.~(\ref{eq:jx1}) and (\ref{eq:jem1}), 
\be
\label{eq:jrecGen}
\jv_\text{rec} = e \qeff \, (n_+ \Vv_+ - n_- \Vv_-) \simeq - i \, \frac{e \qeff \, e^{\prime} \, n_\x}{\w \, m_\x} \, \Ev^\prime_\text{em}
~.
\ee
This visible mCP current $j_\text{rec}$ acts as a source of SM EM fields, and thus can resonantly excite the receiver cavity tuned to the same frequency, depositing a total signal power of~\cite{Hill}
\be
\label{eq:PsigGen}
P_\text{sig} \simeq \frac{Q}{\w} \, \frac{\big| \int_\text{rec} d^3 \xv ~ \Ev_\text{rec}^* \cdot \jv_\text{rec} \big|^2}{\int_\text{rec} d^3 \xv ~ |\Ev_\text{rec}|^2}
~,
\ee
where $\Ev_\text{rec} (\xv)$ is the spatial profile of the excited receiver cavity mode and the integrals are performed over the volume of the receiver cavity. Note that we have not included the effect of an associated mCP visible charge density $\rho_\text{rec}$ in \Eq{PsigGen}; although it can produce small irrotational electric fields,  it cannot excite resonant cavity modes, which are purely solenoidal~\cite{condon1941forced,smythe1988static,collin1990field}.

\subsection{Calculation of the Millicharged DM Signal for a Pair of Cylindrical Cavities}

Eqs.~(\ref{eq:EinvGen}), (\ref{eq:jrecGen}), and (\ref{eq:PsigGen}) can be evaluated for any choice of cavity geometry and emitter mode. Here, we evaluate these expressions for two different representative mode choices in a setup involving coaxially-aligned cylindrical cavities of radius $R$ and length $L$ separated by a distance $d \gg R, L$. In particular, we calculate the signal power in the case that both cavities are operated in either the $\text{TM}_{010}$ or $\text{TE}_{011}$ mode. In cylindrical coordinates, the electric fields profiles in either configuration are given by
\be
\label{eq:modes}
\Ev_\text{em,rec} (\xv) = 
E_\text{em,rec} \times
\begin{cases}
J_0 (\alpha_{0} \rho / R) \, \Theta(z) \, \Theta(L-z) \, \zhat & (\text{TM}_{010})
\\
J_1(\alpha_{1} \rho / R) \, \sin{(\pi z / L)} \, \phihat & (\text{TE}_{011})
~,
\end{cases}
\ee
where $\rho \in [0,R]$, $z \in [0,L]$, $E_\text{em,rec}$ controls the normalization of the respective field, and $\alpha_{0} \simeq 2.40$, $\alpha_{1} \simeq 3.83$ are the first zeroes of $J_0$, $J_1$, respectively. In the first line of \Eq{modes}, we have included the Heaviside step functions $\Theta$ enforcing that the EM fields are confined to $z \in [0,L]$ in order to account for the discontinuity in the electric field at the cavity endcaps located at $z = 0$ and $z = L$. The corresponding resonant frequency of either mode is given by $\w = \alpha_0 / R$ and $\w = \sqrt{(\alpha_1 / R)^2 + (\pi/L)^2}$ for $\text{TM}_{010}$ and $\text{TE}_{011}$, respectively. 

Let us begin by evaluating the invisible field generated by mCPs in an emitter cavity driven in the $\text{TM}_{010}$ mode. Taking the emitter cavity to span $z^\prime \in [0, L]$, the relevant factor in the integrand of \Eq{EinvGen} is
\be
\Ev_\text{em} (\xv^\prime) + \frac{1}{\w^2 - \w_p^{\prime \, 2}} \, \grad \grad \cdot \Ev_\text{em} (\xv^\prime) = E_\text{em} \, J_0 (\alpha_0 \rho^\prime / R) \, \zhat \, \Big( 1 + \frac{1}{\w^2 - \w_p^{\prime \, 2}} \, \partial_{z^\prime} \big[ \delta (z^\prime) - \delta (z^\prime-L) \big] \Big)
~,
\ee
where the factor involving delta functions in the expression above arises from the step functions in \Eq{modes} and effectively incorporates the presence of the mCP density $\rho^\prime$ generated in the emitter cavity near the endcaps. Using this in  \Eq{EinvGen}, the invisible field sourced by mCPs in the emitter cavity is approximately
\be
\label{eq:EprimeTM}
\Ev_\text{em}^\prime (\xv, t) \simeq \frac{e \qeff \, e^\prime \, n_\x}{\alpha_0 \, m_\x} ~ \frac{e^{-i k z} \, (1 - e^{i k L})}{(k z)^2} ~  \bar{E}_\text{em} \, R^2 \, e^{i \w t} \, \zhat
~,
\ee
in the far-field limit $z \gg R, L$, where $\bar{E}_\text{em}=E_\text{em} J_1(\alpha_0)$ is the RMS volume-averaged emitter field and we assumed that $k > 0$ (corresponding to $\w_p^\prime < \w$). Using the above expression in Eqs.~(\ref{eq:jrecGen}) and (\ref{eq:PsigGen}) yields the signal power deposited into the receiver cavity placed a distance $d \gg R, L$ away,
\be
\label{eq:PsigTM}
P_\text{sig} (\text{TM}_{010}) \simeq 64 \pi \, \Big( \frac{e \qeff \, e^\prime \, n_\x}{\alpha_0 \, m_\x} \Big)^4 \, \frac{Q \, \bar{E}_\text{em}^2 \, R^6}{d^4 \, L \, \w^3 \, k^6} \, \sin^4{(k L / 2)}
~.
\ee

Instead, for cavities operated in the $\text{TE}_{011}$ mode, $\grad \cdot \Ev_\text{em} = 0$, such that the second term in the integrand of \Eq{EinvGen} can be dropped. In the far-field limit, we find that the invisible field sourced by mCPs in the emitter cavity is approximately
\be
\label{eq:EprimeTE}
\Ev_\text{em}^\prime (\xv, t) \simeq \frac{\pi^2 \, H_1 (\alpha_1)}{2 \sqrt{2} \, \alpha_1} \, \frac{e \qeff \, e^\prime \, n_\x}{m_\x} ~ \frac{e^{-i k z} \, (1 + e^{i k L})}{z} ~  \bar{E}_\text{em} \, \frac{R^2 \, L}{\pi^2 - k^2 \, L^2} \, e^{i \w t} \, \phihat
~,
\ee
where $\bar{E}_\text{em} = E_\text{em} \, |J_0(\alpha_1)| \, / \sqrt{2}$ is the $\text{TE}_{011}$ RMS volume-averaged emitter field. Comparing Eqs.~(\ref{eq:EprimeTM}) and (\ref{eq:EprimeTE}), we see that the $\text{TE}_{011}$ configuration leads to enhanced fields in the far-field limit, since $E_\text{em}^\prime \propto 1/z$, instead of $E_\text{em}^\prime \propto 1/z^2$ as in the $\text{TM}_{010}$ configuration. For the $\text{TE}_{011}$ driven mode, the corresponding signal is then evaluated to be
\be
\label{eq:PsigTE}
P_\text{sig} (\text{TE}_{011}) \simeq 4 \pi^9 \, \Big( \frac{e \qeff \, e^\prime \, n_\x}{\alpha_1 \, m_\x} \Big)^4 \, \frac{Q \, \bar{E}_\text{em}^2 \, R^6 \, L^3}{d^2 \, \w^3 \, (\pi^2 - k^2 \, L^2)^4} \, H_1^4(\alpha_1) \, \cos^4{(k L / 2)}
~.
\ee
The enhancement of the $\text{TE}_{011}$ configuration in the far-field limit is to be expected. Just as in normal electromagnetism,  if the dark photon is massless then $\Ev_\text{em}^\prime$ is dominantly emitted in directions transverse to the direction of the oscillating dark current $\jv_\text{em}^\prime$, which in turn is aligned with the polarization of the driven emitter's electric field $\Ev_\text{em}$. Thus, the ``longitudinal" $\text{TM}_{010}$ configuration has suppressed sensitivity in the far-field limit. However, as we show below, since $d \sim 1 / \w$ in typical RF LSW experiments, this amounts to only a small penalty in sensitivity, and as a result both setups have comparable sensitivity to the mCDM coupling $\qeff$. 

In our estimates above, we have only investigated coaxially-aligned cavities, finding that the signal in a TE configuration is enhanced compared to that of a TM setup in the far-field limit. We expect similar conclusions to hold for different choices of the cavity alignment. For instance, in the alternative case that the cavities are instead kept parallel and separated in the radial direction, then TM modes involve electric fields which are purely transverse (with respect to the axis connecting the two cavities), while TE modes possess both transverse and longitudinal components. Hence, applying the lessons learned from the above examples, we would expect both mode choices to possess optimal scaling in the far-field limit. Indeed, an explicit calculation confirms this intuition, although we do not present the detailed analysis here.

As discussed near \Eq{EsigParametric}, the signal power scales as $P_\text{sig} \propto (e \qeff \, e^\prime \, n_\x / m_\x)^4 = \eps^4 \, \w_p^{\prime \, 8}$. Interpreting the dark plasma frequency $\w_p^\prime$ as an effective dark photon mass, we see that this result is analogous to direct signals of transverse massive dark photons in the absence of an mCP background, shown in \Eq{massive}. In such searches, the unoptimized ``transverse configuration" scales as $P_\text{sig} \propto \eps^4 \, \mAp^8$, whereas the optimized ``longitudinal configuration" scales instead as $P_\text{sig} \propto \eps^4 \, \mAp^4$. However, this analogy is imperfect, as $\w_p^\prime$ does not constitute a true mass for the dark photon. Indeed, neither of the mCP signals scales as $P_\text{sig} \propto \eps^4 \w_p^{\prime \, 4}$. The origin of this difference is that the plasma frequency and a true mass both modify the dispersion relation of the transverse mode in a similar manner, i.e., $(\w^2 - k^2) \, \Ap_T = (\mAp^2 + \w_p^{\prime \, 2}) \, \Ap_T$, whereas the modifications are entirely different for the longitudinal mode when $k \neq 0$, $(\w^2 - k^2) \, \Ap_L = \big[ \mAp^2 + (1-k^2 / \w^2) \, \w_p^{\prime \, 2} \big] \, \Ap_L$.

\section{Existing Constraints and Future Sensitivity}\label{sec:results}

The expression for the signal power in Eqs.~(\ref{eq:PsigTM}) and (\ref{eq:PsigTE}) can be used to estimate existing limits and future sensitivity of LSW experiments to a terrestrial density of mCDM. We begin by recasting the result of the Dark SRF experiment, which set its first limits in a recent pathfinder run~\cite{Romanenko:2023irv}. Dark SRF employs two coaxial elliptical cavities operated in the $\text{TM}_{010}$ mode. In our calculation, we approximate these cavities as cylinders with radius $R \simeq 10 \ \text{cm}$ and length $L \simeq 5 \ \text{cm}$, separated by a distance of $d \simeq 50 \ \text{cm}$, with the emitter cavity driven with an RMS-averaged field of $\bar{E}_\text{em} \simeq 6 \ \text{MV} / \text{m}$, and the receiver cavity possessing a loaded quality factor of $Q \simeq 3 \times 10^{10}$. This pathfinder run placed a limit on the signal power in the receiver cavity of $P_\text{sig} \simeq 2.5 \times 10^{-16} \ \text{W}$. From \Eq{PsigTM}, we estimate that this places an existing limit of
\be
\label{eq:DarkSRFlimit}
\qeff \lesssim 0.1 \times \bigg( \frac{1}{e^\prime} \bigg) \, \bigg( \frac{m_\x}{1 \ \text{GeV}} \bigg) \bigg( \frac{10^6 \ \text{cm}^{-3}}{n_\x} \bigg)
~.
\ee

Future runs of the Dark SRF experiment will strengthen this sensitivity. The most notable planned improvements are increased frequency matching between the emitter and receiver cavities (a mismatched frequency in the pathfinder run limited the attainable signal power by roughly five orders of magnitude), larger quality factors, and reduced noise. To estimate the reach of a future experiment, we take the same experimental parameters as assumed above, except we assume an average field of $\bar{E}_\text{em} \simeq 25 \ \text{MV} / \text{m}$, a quality factor of $Q = 10^{12}$, and take the limiting noise to be arising from thermal occupation of the EM modes. In this case, if the emitter field's phase is actively monitored, an optimized signal analysis leads to an effective noise power of $P_\text{noise} \simeq T / t_\text{int}$~\cite{Graham:2014sha}, where $T = 10 \ \text{mK}$ is the temperature of the receiver cavity, and $t_\text{int} = 1 \ \text{yr}$ is the total experimental integration time. Setting $P_\text{sig} = P_\text{noise}$, we estimate a future sensitivity of
\be
\label{eq:futureDarkSRF}
\qeff \lesssim \bigg( \frac{1}{e^\prime} \bigg) \, \bigg( \frac{m_\x}{1 \ \text{GeV}} \bigg) \bigg( \frac{10^6 \ \text{cm}^{-3}}{n_\x} \bigg) 
\times
\begin{cases}
2 \times 10^{-6} \ (\text{TM}_{010})
\\
6 \times 10^{-7} \ (\text{TE}_{011})
~.
\end{cases}
\ee
In the first and second lines in the expression above, we have assumed that the cavities are operated in the $\text{TM}_{010}$ or $\text{TE}_{011}$ mode, respectively, which shows that a TE configuration results in an $\order{1}$ enhanced sensitivity to the mCP coupling. We also note that the CROWS experiment operated in the optimized configuration, employing coaxially-aligned cavities in the $\text{TE}_{011}$ mode~\cite{Betz:2013dza}. However, this does not result in a competitive limit compared to that derived from Dark SRF in \Eq{DarkSRFlimit}, since the mild enhancement due to the TE mode is not sufficient to overcome the penalty of the smaller quality factor and field strength.  

\begin{figure}[t]
\centering
\includegraphics[width = 0.5\textwidth]{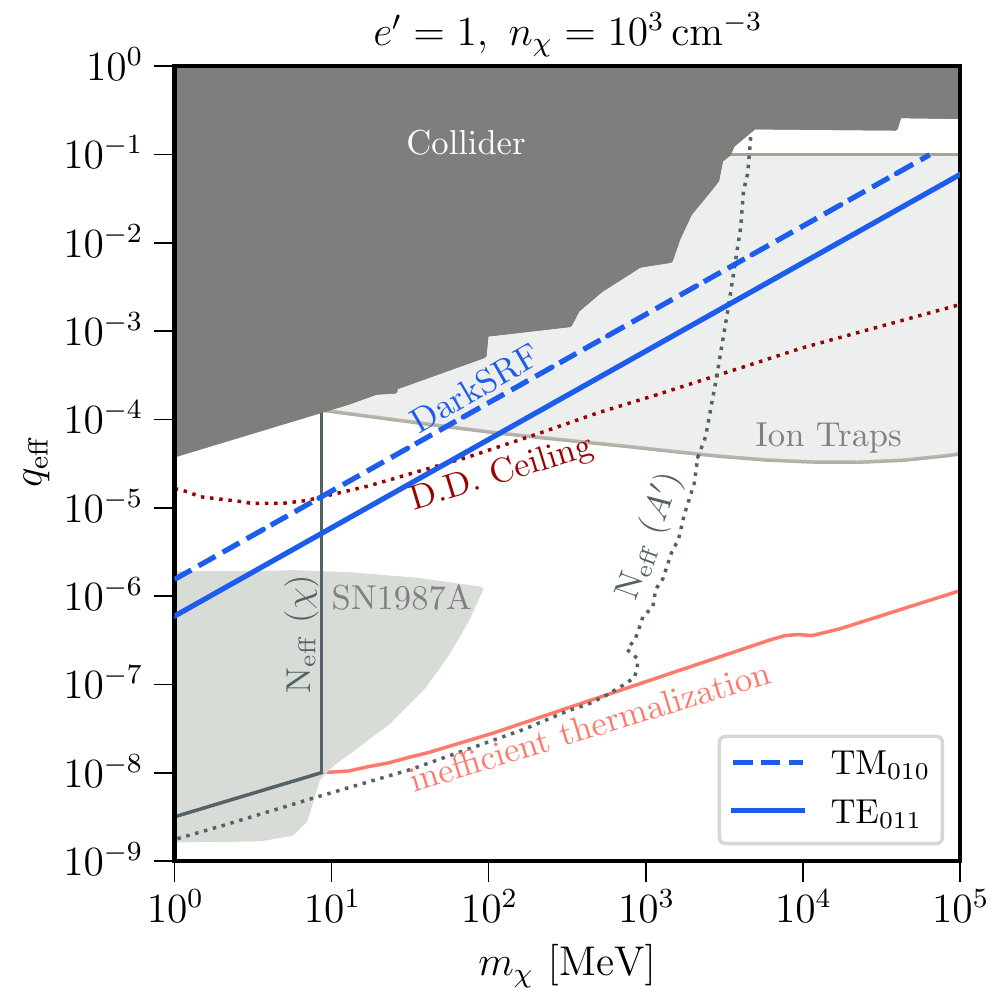}~\includegraphics[width = 0.5\textwidth]{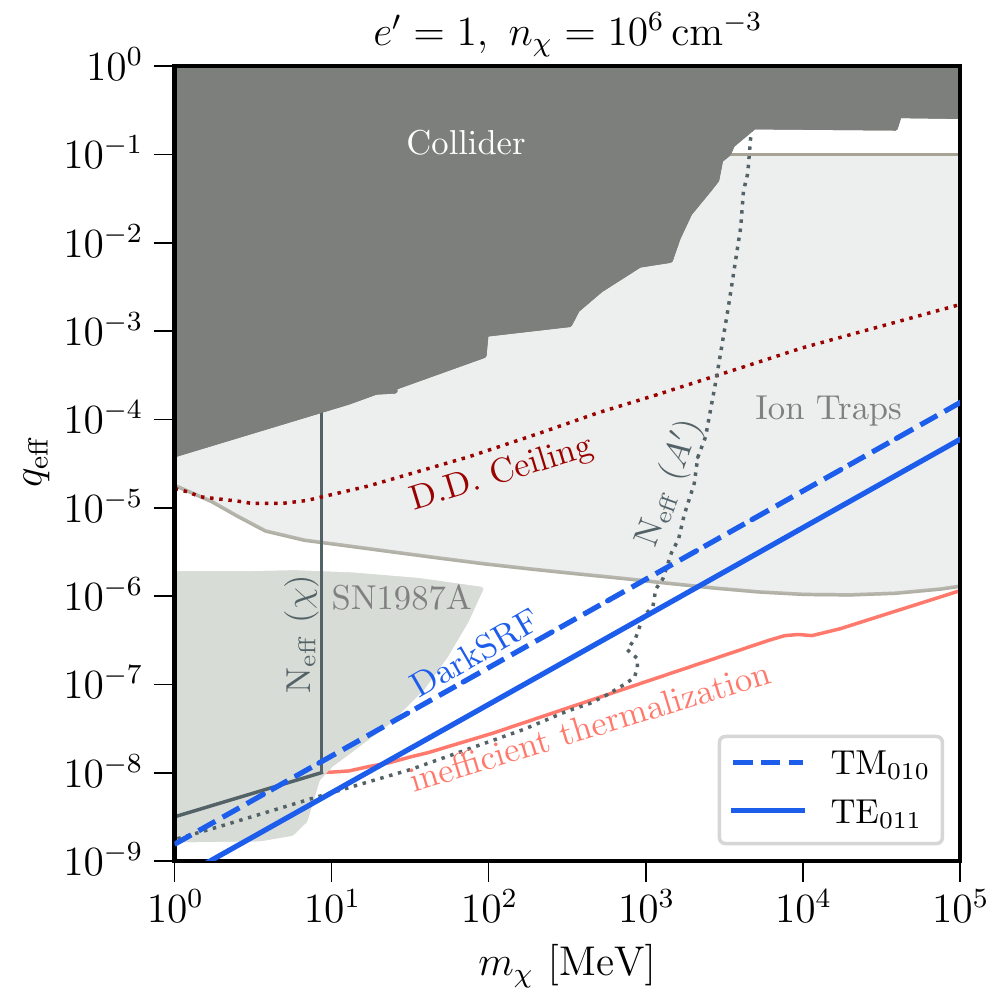}
\caption{The future sensitivity (blue lines), as in \Eq{futureDarkSRF}, of Dark SRF to mCPs with a laboratory number density of $n_\chi = 10^3\ \text{cm}^{-3}$ (left) or $n_\chi = 10^6\ \text{cm}^{-3}$ (right), fixing $e'=1$ and taking the cavities to be operated in the optimal $\text{TE}_{011}$ (solid) or the sub-optimal $\text{TM}_{010}$ (dashed) mode-configuration. Limits from collider searches and SN1987A are shown as shaded grey regions. Model-dependent bounds from modifications to the cosmological expansion rate are shown as solid and dotted gray lines. Limits from ion traps~\cite{Budker:2021quh} for $n_\chi = 10^3\ \text{cm}^{-3}$ ($10^6\ \text{cm}^{-3}$) are also shown in the left (right) panel. Note that ion traps and Dark SRF are not sensitive to $\qeff\gtrsim 0.1$, since in this case mCPs are not able to penetrate typical laboratory devices~\cite{Budker:2021quh}. Above the dotted dark red line, mCP relics rapidly thermalize in Earth's atmosphere and crust before encountering terrestrial direct detection experiments. Below the solid light red line, $\qeff$ is not sufficiently large for mCPs to thermalize within any region of the Earth.}
\label{fig:ReachPlot1e3}
\end{figure}

\begin{figure}[t]
\centering
\includegraphics[width = 0.5\textwidth]{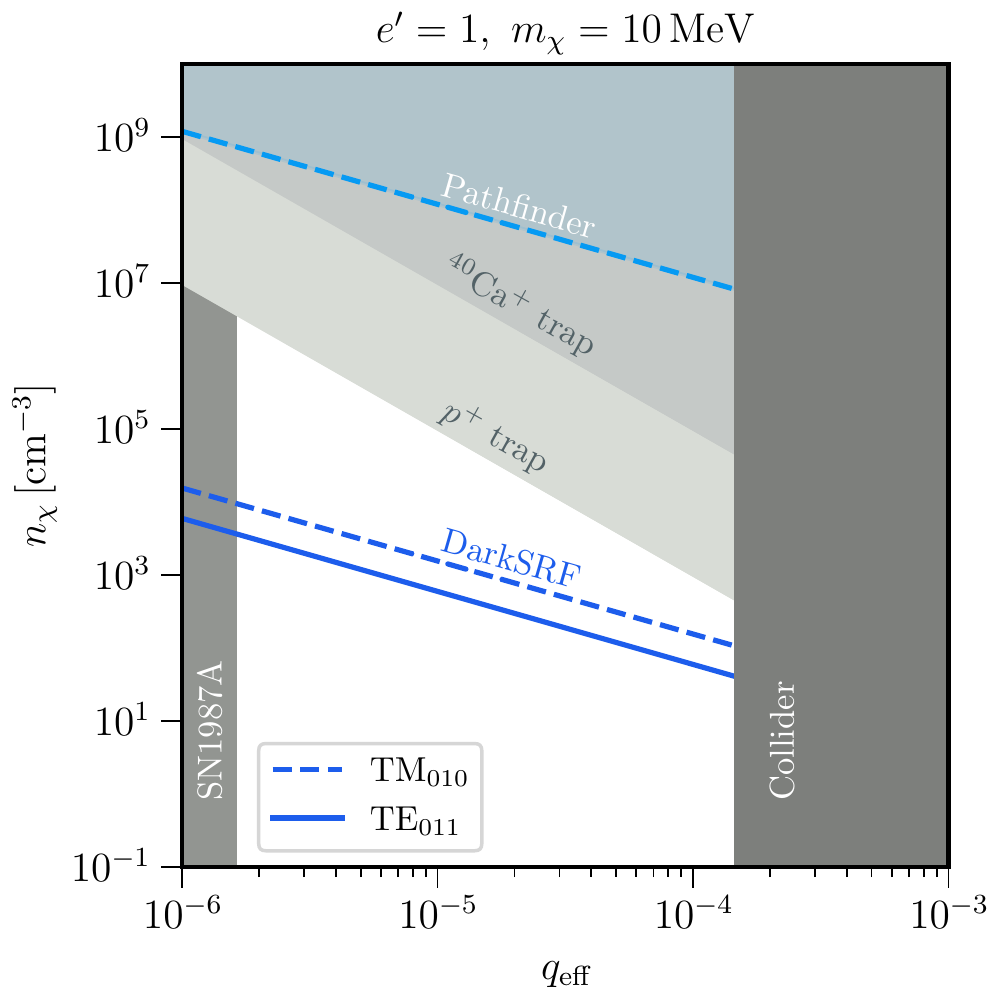}~\includegraphics[width = 0.5\textwidth]{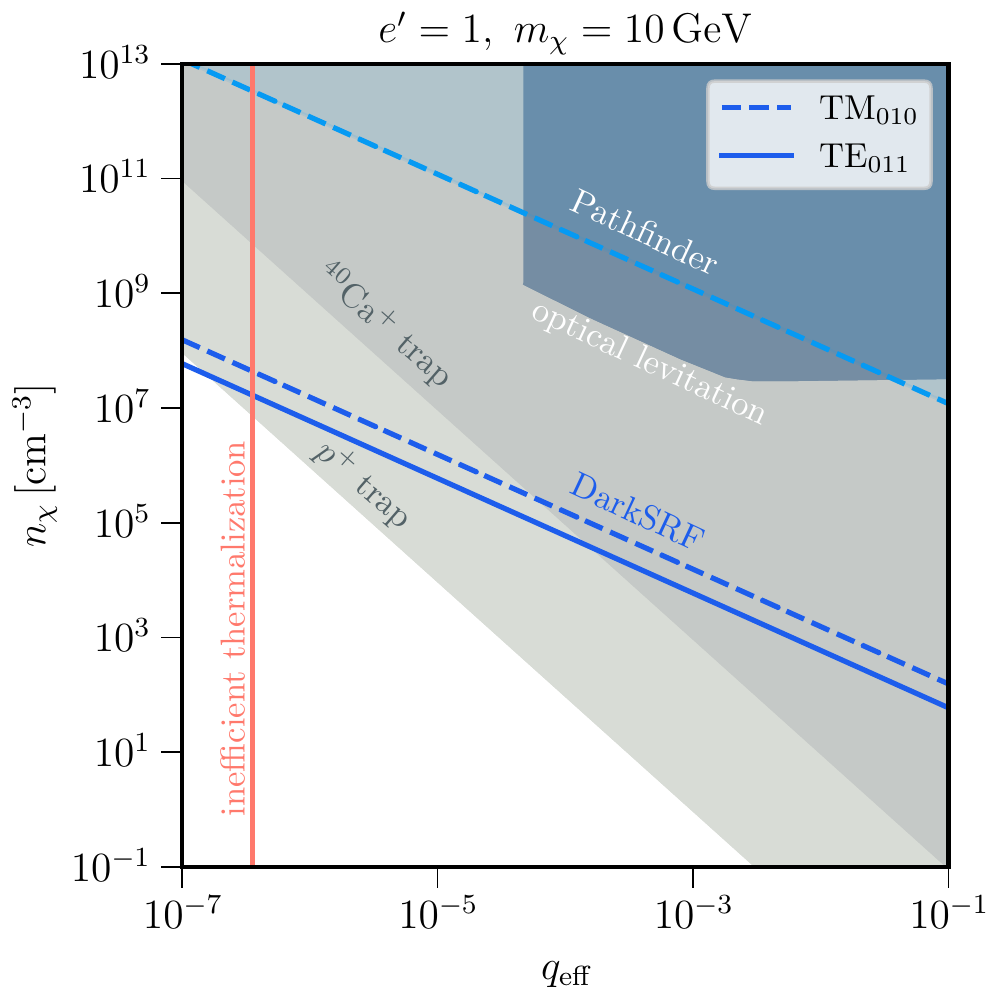}
\caption{As in Fig.~\ref{fig:ReachPlot1e3}, but fixing $m_\chi = 10\ \text{MeV}$ (left panel) or $m_\chi = 10 \ \text{GeV}$ (right panel) in the parameter space spanned by $n_\x$ and $\qeff$. Dark SRF projections are shown as blue lines. Compared to Fig.~\ref{fig:ReachPlot1e3}, we also show existing limits recast from the recent Dark SRF pathfinder run (light blue)~\cite{Romanenko:2023irv} as well as optically-levitated nanospheres~\cite{Afek:2020lek}.}
\label{fig:ReachPlotNX}
\end{figure}

The projected sensitivity of Dark SRF is shown as solid (dashed) blue lines in Figs.~\ref{fig:ReachPlot1e3} and \ref{fig:ReachPlotNX}, assuming an experimental setup consistent with the second (first) line of \Eq{futureDarkSRF}. In Fig.~\ref{fig:ReachPlot1e3}, we illustrate the reach in $\qeff$ as a function of the mCP mass $m_\x$, fixing the dark gauge coupling to $e^\prime = 1$ and the ambient mCP density within the cavities to either $n_\x  = 10^3 \ \text{cm}^{-3}$ (left panel) or $n_\x  = 10^6 \ \text{cm}^{-3}$ (right panel). As discussed in Section~\ref{sec:milliQ}, this density can be parametrically different from the galactic density and average mCP density near Earth's surface, since $n_\x$ depends both on the density and temperature of the surrounding environment~\cite{Berlin:2023zpn}. We discuss this in more detail below. 
Note that in our projections, we have adopted a noise temperature of $T = 10 \ \text{mK}$, which corresponds to the temperature of the receiver cavity; the emitter cavity, which contains large driven fields, can only be feasibly cooled to $T \sim 1 \ \text{K}$. Hence, our adopted values of $n_\x$ should be interpreted as the geometric mean of the mCP density in both cavities. Also shown in solid gray are existing limits from accelerators~\cite{Prinz:1998ua,ArgoNeuT:2019ckq,milliQan:2021lne,Davidson:2000hf,ArguellesDelgado:2021lek}, observations of SN1987A~\cite{Chang:2018rso}, and anomalous heating in ion traps~\cite{Budker:2021quh}.\footnote{We have chosen to show ion trap limits assuming a comparable fixed density within the trap itself. We thank Hari Ramani for providing the corresponding limits for us here, which have been appropriately modified from those presented in Ref.~\cite{Budker:2021quh}.} In Fig.~\ref{fig:ReachPlotNX}, we instead show the mCP parameter space spanned by the charge $\qeff$ and laboratory density $n_\x$, fixing $e^\prime = 1$ and $m_\x = 10 \ \text{MeV}$ (left panel) or $m_\x = 10 \ \text{GeV}$ (right panel). From Figs.~\ref{fig:ReachPlot1e3} and \ref{fig:ReachPlotNX}, we see that LSW signals of mCPs have a more favorable scaling than other techniques for smaller masses.

Model-dependent cosmological limits are shown as either solid or dotted gray lines in Fig.~\ref{fig:ReachPlot1e3}. Additional dark radiation in the form of mCPs and dark photons may unacceptably alter the expansion rate in the early Universe. This is usually parametrized as an increased effective number of neutrino species $N_\text{eff}$. Within parameter space bounded by the contour labelled ``$N_\text{eff} (\x)$," mCPs efficiently thermalize in the early universe and directly modify the expansion rate during the time of nucleosynthesis~\cite{Creque-Sarbinowski:2019mcm}. Within the dotted contour labelled ``$N_\text{eff} (\Ap)$," mCPs can indirectly lead to observable modifications to $N_\text{eff}$, since upon thermalizing with the SM bath they generate an appreciable density in dark photon radiation via $\x \x \to \Ap \Ap$~\cite{Vogel:2013raa,Adshead:2022ovo}. Such bounds are weakened for $m_\x \gtrsim 1 \ \text{GeV}$, in which case the relative contribution of the $\Ap$ density is reduced once the SM decouples from the dark sector at temperatures $T \lesssim m_\x$, before numerous mass thresholds in the SM heat the visible sector compared to the dark radiation. However, these limits are alleviated in non-standard cosmologies. For instance, reheat temperatures of the universe as small as $T_\text{reheat} \sim 5 \ \text{MeV}$ are compatible with standard nucleosynthesis~\cite{deSalas:2015glj,Hannestad:2004px,Kawasaki:2000en}, yet can be invoked to prevent the thermalization of mCPs heavier than this scale. In this case, non-thermal mechanisms can give rise to a substantial density of mCPs with $m_\x \gtrsim 10 \ \text{MeV}$, alleviating the bounds labelled as ``$N_\text{eff} (\Ap)$." While low reheat temperatures do not prevent the thermalization of sub-MeV mCPs, alleviating the bounds labelled ``$N_\text{eff} (\x)$" is possible for models in which the millicharge sector's coupling to SM matter is suppressed at large densities and early cosmological times (see, e.g., Refs.~\cite{Masso:2006gc,DeRocco:2020xdt}).

Above the dotted dark red line labelled ``D.D. ceiling" in Fig.~\ref{fig:ReachPlot1e3}, mCP relics rapidly thermalize in Earth's atmosphere and crust before encountering terrestrial direct detection experiments searching for DM-SM scattering~\cite{Emken:2019tni}. Hence, above this contour, scattering-based direct detection searches are not sensitive to such strongly-coupled DM subcomponents, since the particles present in such detectors do not have sufficient energy to scatter and deposit signals above threshold. Below this line, the sensitivity of these experiments depends on the particular value of the galactic mCP density $n_\x^{(\text{gal})}$, which can be orders of magnitude smaller than the fixed laboratory density $n_\x$. For instance, for $1 \ \text{GeV} \lesssim m_\x \lesssim 100 \ \text{GeV}$, laboratory overdensities  in cryogenic detectors can be as large as $n_\x / n_\x^{(\text{gal})} \sim 10^{18}$~\cite{Berlin:2023zpn}. Using the formalism developed in Ref.~\cite{Berlin:2023zpn} to translate between $n_\x$ and $n_\x^{(\text{gal})}$ and the various direct detection limits presented in Ref.~\cite{Emken:2019tni}, we find that direct detection experiments can place limits on minimal models in the parameter space below the red dotted line in Fig.~\ref{fig:ReachPlot1e3} for $m_\x \lesssim 1 \ \text{GeV}$ and $m_\x \gtrsim 100 \ \text{GeV}$, corresponding to where $n_\x^{(\text{gal})}$ is typically only a few orders of magnitude smaller than $n_\x$. 
However, we note that such bounds are alleviated for $m_\x \lesssim 1 \ \text{GeV}$ in non-minimal models where mCPs weakly couple to an additional long-ranged force, resulting in larger terrestrial overdensities for sub-GeV masses~\cite{Acevedo:2023owd}. Finally, below the solid light red line labelled ``inefficient thermalization" in Fig.~\ref{fig:ReachPlot1e3}, $\qeff$ is not sufficiently large for mCPs to thermalize within any region of the Earth~\cite{Berlin:2023zpn}; thus, in assuming large terrestrial overdensities, we are restricted to consider couplings above this contour. 

In Fig.~\ref{fig:ReachPlotNX} we only show the constraints/projections for Dark SRF up to values of $n_\chi$ where $\w_p^\prime \sim \w / 10$. From the discussion in Section~\ref{sec:signal}, it is evident that when $\w_p^\prime$ approaches $\w$, some of the simplifications we have made no longer apply. We also see from \Eq{EinvGen} that when $\w_p^\prime \gtrsim \w$, the invisible dark field produced in the emitter cavity will be exponentially suppressed at distances $d\gtrsim 1/\w_p^\prime$. As a result, the Dark SRF sensitivity will rapidly deteriorate for higher values of $n_\chi$ inside either cavity.

\section{Conclusion}
\label{sec:conclusion}

Additional $U(1)^\prime$ gauge groups that are kinetically-mixed with the visible photon are motivated both by ultraviolet completions of the Standard Model as well as by their ability to provide a natural dark matter candidate. Dark states that are directly charged under this additional gauge group would appear as millicharged particles (mCPs) under electromagnetism. It has previously been shown that if their effective charge is sufficiently large, the thermalized terrestrial abundance of such mCPs can be far greater than the average galactic density, but would have gone undetected by standard direct detection experiments, motivating alternative experimental techniques.

Although direct signals of dark photons decouple in the massless limit, indirect signals may still arise in an mCP background. In particular, we have shown that the light-shining-through-wall (LSW) Dark SRF experiment at FNAL~\cite{Romanenko:2023irv}, operated to search for massive dark photons, also has sensitivity to similar signals of massless dark photons arising from EM-induced disturbances of mCP dark matter subcomponents. Owing to the nature of the signal, the optimal geometry for the fields in the emitter and receiver cavities is one in which the electric fields in both cavities are transverse. This is unlike typical LSW searches for massive dark photons in which case the longitudinal arrangement is optimal. However, although the longitudinal configuration has suppressed sensitivity to mCPs in the far-field limit, the typical separation of the cavities in LSW experiments is such that this only results in an $\order{1}$ penalty in sensitivity to the mCP charge $\qeff$. The scaling of the experimental sensitivity is such that future runs of the Dark SRF experiment can set the best limits on mCPs in certain regions of parameter space, particularly at low masses and small terrestrial abundances.

\acknowledgements

We thank Hari Ramani for providing the appropriately re-scaled constraints on mCPs from ion traps appearing in Fig.~\ref{fig:ReachPlot1e3}. This  material  is  based  upon  work  supported  by  the U.S.\ Department of Energy,  Office of Science,  National Quantum  Information  Science  Research  Centers,   Superconducting  Quantum  Materials  and  Systems  Center (SQMS) under contract number DE-AC02-07CH11359. Fermilab is operated by the Fermi Research Alliance, LLC under Contract DE-AC02-07CH11359 with the U.S.\ Department of Energy. The work of SARE was supported by SNF Ambizione grant PZ00P2\_193322, \textit{New frontiers from sub-eV to super-TeV}. The work of JIR is supported by the Natural Sciences and Engineering Research Council (NSERC) of Canada. Research at Perimeter Institute is supported in part
by the Government of Canada through the Department of Innovation, Science and Economic
Development Canada and by the Province of Ontario through the Ministry of Colleges and
Universities.

\bibliography{bibliography}

\end{document}